\documentclass[]{spie}  
\usepackage[]{graphicx}
\usepackage{units}
\usepackage{wrapfig}
\usepackage{subfigure}
\usepackage{upgreek}
\usepackage{amsmath}
\usepackage[latin1]{inputenc}
\usepackage[T1]{fontenc}
\usepackage{amsmath}
\usepackage{url}
\usepackage{float}
\usepackage{caption}

\def\trademark{{\ooalign{\hfil\raise.07ex\hbox{R}\hfil\crcr\mathhexbox20D}}}

\title{The Actuator Design and the Experimental Tests
of a New Technology Large Deformable Mirror
for Visible Wavelengths Adaptive Optics} 

\author{
Ciro~Del~Vecchio\supit{a}
Guido~Agapito\supit{a}
Carmelo~Arcidiacono\supit{a}
Luca~Carbonaro\supit{a}
Fabrizio~Marignetti\supit{b}
Enzo~De~Santis\supit{b}
Valdemaro~Biliotti\supit{a}
and
Armando~Riccardi\supit{a}
\skiplinehalf
\supit{a}INAF-OAA, Largo E. Fermi 5, I-50125 Firenze, Italy\\
\supit{b}DIEI~--~Università di Cassino e dell'Alto Lazio, Via G.\ Di Biasio 43, I-03043 Cassino, Italy\\
}

\authorinfo{Further author information: (Send correspondence to C. Del Vecchio)\\
C. Del Vecchio: E-mail: cdelvecchio@arcetri.astro.it, Telephone: +39 055 2752261\\
G. Agapito: E-mail: agapito@arcetri.astro.it, Telephone: +39 055 2752315\\
C. Arcidiacono: E-mail: carmelo@arcetri.astro.it, Telephone: +39 055 2752293 \\
L. Carbonaro: E-mail: carbo@arcetri.astro.it, Telephone: +39 055 2752314\\
F. Marignetti: E-mail: marignetti@unicas.it, Telephone: +39 0776 2993716\\
E. De Santis: E-mail: depotenzo@gmail.com, Telephone: +39 0776 2994705\\
V. Biliotti: E-mail: biliotti@arcetri.astro.it, Telephone: +39 055 2752295\\
A. Riccardi: E-mail: ariccardi@arcetri.astro.it, Telephone: +39 055 2752207
}

\makeatletter
	\newcommand\figcaption{\def\@captype{figure}\caption}
	\newcommand\tabcaption{\def\@captype{table}\caption}
\makeatother

\begin{document} 
\maketitle 
\pdfoutput=1
\begin{abstract} 
	Recently, Adaptive Secondary Mirrors showed excellent on-sky
	results in the Near Infrared wavelengths. They currently provide
	30mm inter-actuator spacing and about 1 kHz bandwidth. Pushing
	these devices to be operated at visible wavelengths is a
	challenging task. Compared to the current systems, working in
	the infrared, the more demanding requirements are the higher
	spatial resolution and the greater correction bandwidth. In fact,
	the turbulence scale is shorter and the parameter variation is
	faster. Typically, the former is not larger than 25 mm (projected
	on the secondary mirror) and the latter is 2 kHz, therefore
	the actuator has to be more slender and faster than the current
	ones. With a soft magnetic composite core, a dual-stator and
	a single-mover, VRALA, the actuator discussed in this paper,
	attains unprecedented performances with a negligible thermal
	impact. Pre-shaping the current required to deliver a given stroke
	greatly simplifies the control system, whose output supplies
	the current generator. As the inductance depends on the mover
	position, the electronics of this generator, provided with an
	inductance measure circuit, works also as a displacement sensor,
	supplying the control system with an accurate feed-back signal.
	A preliminary prototype, built according to 
	the several FEA thermo-magnetic analyses, has undergone
	some preliminary laboratory tests. The results of these checks, matching
	the design results in terms of power and force, show that the
	the magnetic design addresses the severe specifications.
\end{abstract}




\section{THE SCIENTIFIC AND TECHNOLOGICAL RATIONALE}					\label{sec:rationale}

The Extremely Large Telescopes (ELT) phase A
studies\cite{TMT_MARS08_1,GMT_MARS08_1,gilmozzi_2007}
are disclosing the design of future state-of-the-art ground
based optical and NIR astronomical facilities.
The role of current 8-10 meter class telescope will be revised
in order to matches new science and new ELT discoveries. A way
to obtain possible and qualified synergies is to work at similar
spatial resolution. Given the telescopes resolutions and
the extensive use of Adaptive Optics (AO) techniques on both classes,
we propose to consider AO at visible wavelengths for 8 meter
telescopes in order to obtain similar wavelength on diameter
ratios. The AO@SW (Adaptive Optics at Short Wavelengths) 
proposal investigates this possibility through numerical
simulations\cite{ADOPT_AMS12_2}.

One of the main science drivers for ELTs is the star formation
histories of galaxies through stellar population studies\cite{hook}.
In particular, resolved stellar population analysis in regions not
accessible today because of extreme crowding (because of the spatial resolution limit
of the instruments) will allow to analyze the
formation histories of our portion of the universe up to Virgo or
(hopefully) Coma galaxies clusters\cite{deep_2011}. A further, interesting
science drive is the study the populations of our and close by
galaxies of single age stars ensembles such as Globular and Open Cluster
looking at magnitude fainter than turn-off point.

It is well known that large observed wavelengths boosts the sensitivity
of the color magnitude diagram (CMD) to the age and to the metallicity
of the star clusters. At ELT spatial resolutions this corresponds to NIR
CMD such K vs J-K: however at the same high spatial resolution an 8 meter
class telescope may provide the photometry measurements of the same stars
detected by the ELT but at the V or R band offering a much larger baseline
for the CMD.

\section{INTRODUCTION}					\label{sec:intro}

The current Adaptive Secondary Mirrors (ASM),
with a $\unit[30]{mm}$ inter-actuator spacing and a $\unit[1]{kHz}$
bandwidth, showed recently excellent on-sky results at NIR
wavelengths. Operating an ASM at visible wavelengths is a challenging
task: the more demanding requirements are the higher spatial resolution
and the greater correction bandwidth. In fact, the turbulence scale
is shorter and the parameter variation is faster. Typically, the
former is not larger than $\unit[25]{mm}$ (projected on the secondary mirror)
and the latter is $\unit[2]{kHz}$.
As a consequence, the actuator for a visible wavelength AO system has to be more
slender and faster than the current ones.
The actuators developed for AO
at NIR wavelengths of the 8 meter class telescopes,
described in [\citenum{LBT_VENICE01}], are not suitable for shorter
wavelengths~---~where a low-order and long-stroke ASM
is required. Therefore, different designs, such as the ones
described in [\citenum{TLIPMM_2006,ADOPT_MARS08_1}], have
been exploited, among the large variety of linear motors developed
in the past years\cite{ieee_2010_1,eerr_2009_1,ieee_2010_2}.
Focusing the magnetic flux density by
means of suitable statoric yokes-like components of soft iron
material allowed to decrease by one order of magnitude the power
dissipated to actuate the correction force. Nevertheless, those
very good performances were obtained with a significant mechanical
complexity.

With a soft magnetic composite core consisting of a dual-stator
and a single-mover, VRALA\cite{ADOPT_SD10_1,2010_comsol}
(Variable Reluctance Adaptive mirror Linear Actuator),
the actuator discussed in this paper, attains unprecedented
performances with a negligible thermal impact.
With a simpler geometry an even better performances, it is the
ideal candidate for the AO actuators at visible wavelengths.
Pre-shaping the
current required to deliver a given stroke greatly simplifies the
control system, whose output supplies the current generator. As
the inductance depends on the mover position, the electronics
of this generator, provided with an inductance measure circuit,
works also as a displacement sensor, providing the control system
with an accurate feed-back signal. The entire design process
is based on numerical simulations: the Finite Element Analysis
(FEA) method is applied
for the electromagnetic, mechanical and thermal studies, a CAD
electronics design and simulation system defines the electronic
hardware architecture, and the control system is developed by means
of the Matlab/Simulink\textsuperscript{\textregistered} software.
A preliminary prototype, 
built according to this simple, effective and low power consumption
design, is aimed for checking the study results~---~and possibly
correct/revise any critical issue.

Recalling the description given in [\citenum{ADOPT_SD10_1,2010_comsol}],
the basic static component of VRALA is a
cylindrical, hollow shaped soft iron stators, that accommodate the
coils. The flux lines of the magnetic field produced by the current
flowing in the coil are conveyed into a mover, a disk also built
of soft iron and facing the stator, through an air gap. As the
magnetic pressure in that gap works as a pull-only force on the
mover, a second stator, placed symmetrically with respect to the
mover, is needed to produce the pushing force. The force is applied
to the ASM by a shaft fixed to the mover,
mounted in the stator central hole parallelly to its axis by
means of two membranes.
As the current/force transfer function is highly non-linear~---~the
mover position and the magnetic characteristics avoid to
analytically identify such a function~---~the control system design
is based on an open-loop preshaper current function, determined via
a look-up table obtained computing the magnetostatics of the system
for many currents and positions, and on a very simple, proportional-only
closed loop which takes care of the (very small) residual corrections.
The current command signal generated by the control system feeds
a power supplier, whose electronics is designed to deliver the
needed current and to measure the device inductance, which depends
on the mover position, at the same time. Providing the control
system with an accurate feed-back signal, the electronics of the
power supplier works also as a displacement sensor. The simpleness
and efficiency of the electronic design, which adopts the more
recent devices developed for power management applications, such
as switching power suppliers, allows to minimize the thermal
impact in the delicate optical environment.

\section{The Design}		\label{sec:fem}

\begin{figure}[t]
	\begin{minipage}[c]{0.55\textwidth}
		\centering
		\includegraphics[height=1.20\textwidth]{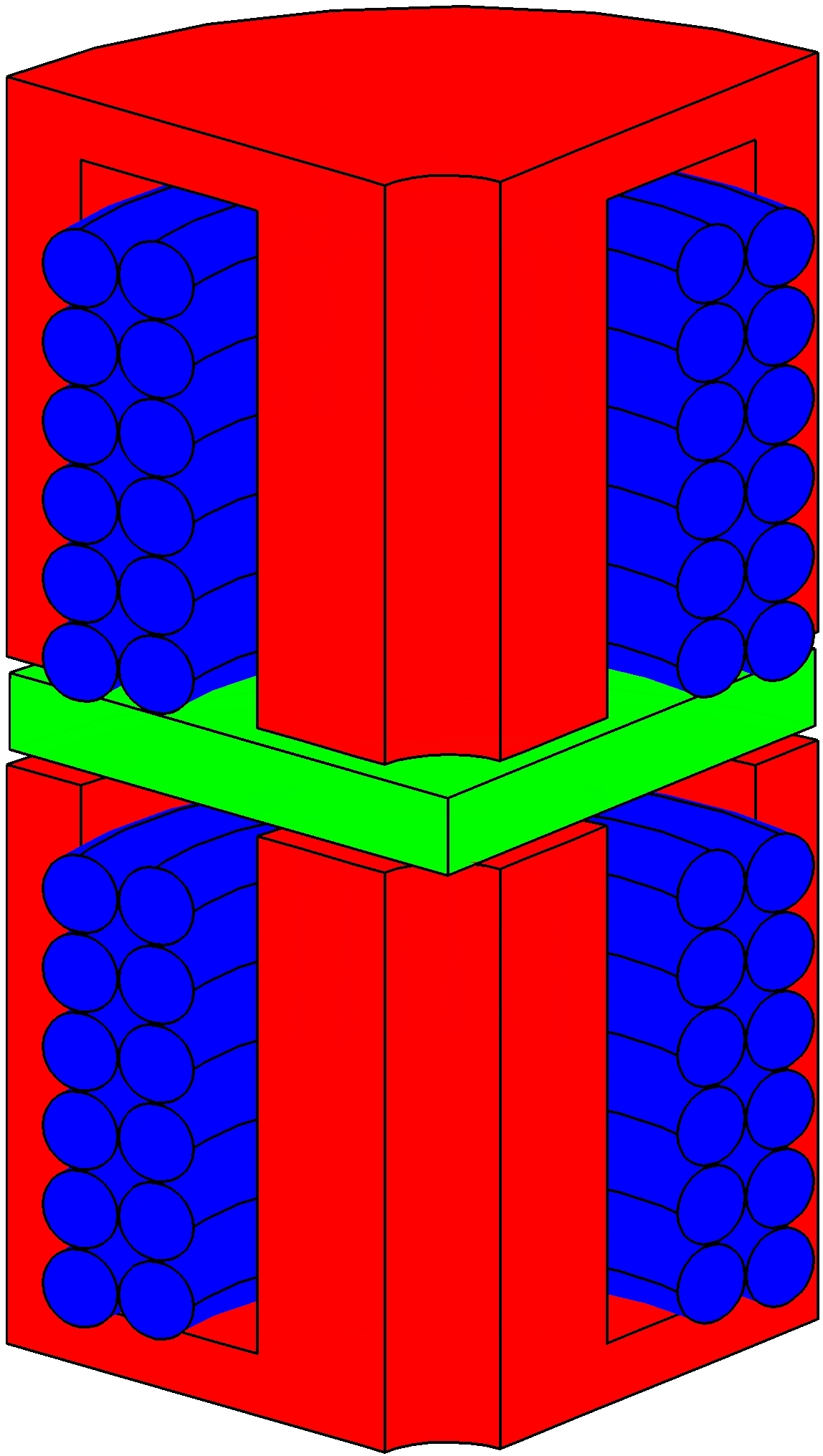}
		\label{fig:3d_schem}
		\captionof{figure}{Schematic view of the actuator.}
	\end{minipage}%
	\begin{minipage}[c]{0.45\textwidth}
		\centering
		\begin{tabular}{|l|r|}
			\hline
			rms force (turbulence correction)		& $\unit[.363]{N}$					\\
			\hline
			max force (static)				& $\unit[.36]{N}$					\\
			\hline
			max force (dynamic)				& $\unit[1.27]{N}$					\\
			\hline
			stroke (usable)					& $\unit[\pm 150]{\upmu m}$				\\
			\hline
			stroke (mechanical)				& $\unit[\pm 200]{\upmu m}$				\\
			\hline
			bandwidth					& $\unit[2]{kHz}$					\\
			\hline
			typical inter-actuator spacing			& $\unit[25]{mm}$					\\
			\hline
			typical actuator length				& $\unit[\leq 60]{mm}$					\\
			\hline
			typical mover mass				& $\unit[\leq 10 ]{g}$					\\
			\hline
		\end{tabular}
		\label{tab:specs}
		\captionof{table}{Schematic view of the actuator.}
	\end{minipage}
\end{figure}

{\mbox{Comsol Multiphysics\textsuperscript{\textregistered}}}, the code
adopted for the numerical analyses\cite{comsol_2011}, allows
to easily generate a FEM mesh both accurate and numerically
smooth, so that exploring the effect of the material choices and
the geometry variations on the solutions is quite fast and accurate.
A single Matlab script can accomplish in
a single process such a geometrical task, the geometry mesh, the embedding
of the mesh elements in the electromagnetic
``azimuthal currents'' module, taking into account the
physical properties of the chosen materials, including the air, the
non linear solution of the non linear system
for the magnetic potential variable, and the post-processing
computation of the magnetic force.
This process is the same described in [\citenum{ADOPT_SD10_1}], with
two relevant differences. First, the {\em{multi-turn}} capability in Comsol
allows to apply the coil a current excitation and to avoid meshing the actual
coil. Second, because of the very high current time derivatives to be
applied by the control system, described in {\mbox{Sec.\ \ref{sec:cs}}}, that
induces high eddy currents in both the stators and the mover, these components
have been built with Somaloy. The model geometry is sketched in
{\mbox{Fig.\ \ref{fig:3d_schem}}}; the specifications are summarized
{\mbox{in Tab.\ \ref{tab:specs}}}.

\subsection{MAGNETOSTATICS: COMPUTING THE FORCE}					\label{sec:MS}

\begin{figure}[H]
	\begin{center}
		\includegraphics[width=.99\textwidth]{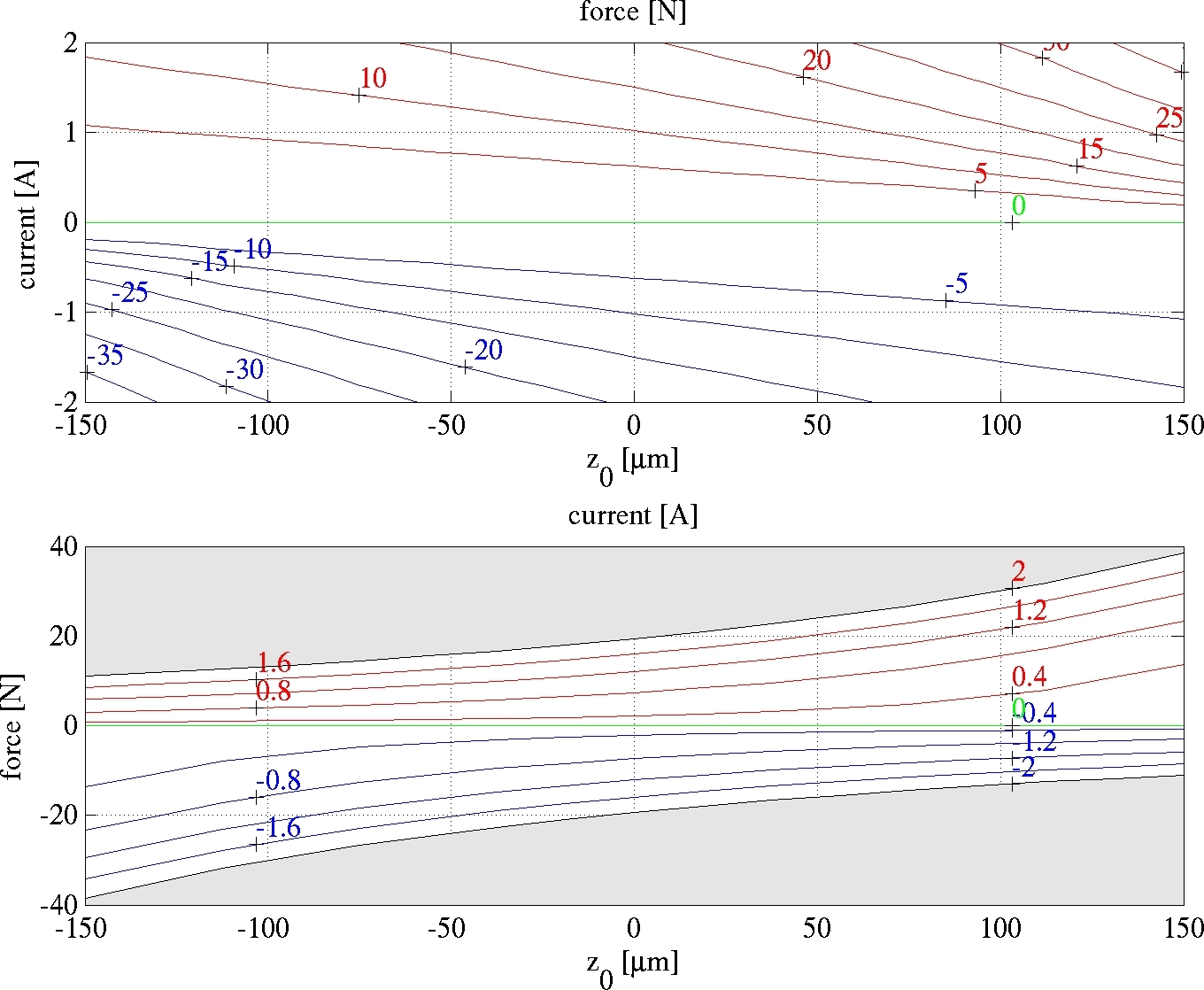}
		\caption{Force $F$ as a function of the mover position
		$z$ and current $I$ (top) and $I$ as a function of $F$
		and $z$ (bottom). Positive values of $I$ correspond 
		to currents flowing in the top coil, whereas negative
		values to currents flowing in the bottom one. Note that
		the shaded areas in the bottom plot indicate undeliverable
		forces.}
		\label{fig:tf}
	\end{center}
\end{figure}

The magnetostatic computations are aimed to define the optimized geometry,
by means of the process discussed in [\citenum{comsol_2011}],
as well as to determine the relationship between the force $F$,
the mover position $z$ and the coil current $I$, needed by the
control system, discussed in {\mbox{Sec.\ \ref{sec:cs}}}. The geometry giving the maximum
efficiency, namely $\epsilon = \unit[4.65]{N \times W^{-1}}$,
is summarized
in {\mbox{Tab.\ \ref{tab:optim}}}; the values of the above mentioned function
are given in {\mbox{Fig.\ \ref{fig:tf}}}, where both $F=f(z,I)$ and its inverse
function $I=g(z,F)$, the one actually needed by the control system, are
plotted.

\begin{table}[H]
	\caption{The physical and geometrical parameters of the optimized geometry.}
	\label{tab:optim}
	\begin{center}
		\begin{tabular}{|c|c||c|c|}
				\hline
				outer radius of stator		&	$\unit[7]{mm}$		&	height of coil slot	&	$\unit[5.9]{mm}$	\\
				\hline
				inner radius of stator		&	$\unit[1]{mm}$		&	width of coil slot	&	$\unit[2.3]{mm}$	\\
				\hline
				height of stator		&	$\unit[7.5]{mm}$	&	gap height		&	$\unit[.2]{mm}$		\\
				\hline
				height of stator slot		&	$\unit[5.9]{mm}$	&	wire outer radius	&	$\unit[120.0]{\upmu m}$	\\
				\hline
				outer radius of mover		&	$\unit[6.95]{mm}$	&	insulation thickness	&	$\unit[10.0]{\upmu m}$	\\
				\hline
				inner radius of mover		&	$\unit[0]{mm}$		&	coil resistance	 	&	$\unit[2.257]{\Omega}$	\\
				\hline
				height of mover			&	$\unit[1]{mm}$		&	number of turns	 	&	$\unit[240]{}$		\\
				\hline
				mean radius of coil slot	&	$\unit[4.62]{mm}$	&	filling factor	 	&	$\unit[.627]{}$		\\
				\hline
		\end{tabular}
	\end{center}
\end{table}

\subsection{FREQUENCY-DOMAIN: MEASURING THE DISPLACEMENT}				\label{sec:freq_analysis}

\begin{figure}[H]
	\begin{center}
		\includegraphics[width=.99\textwidth]{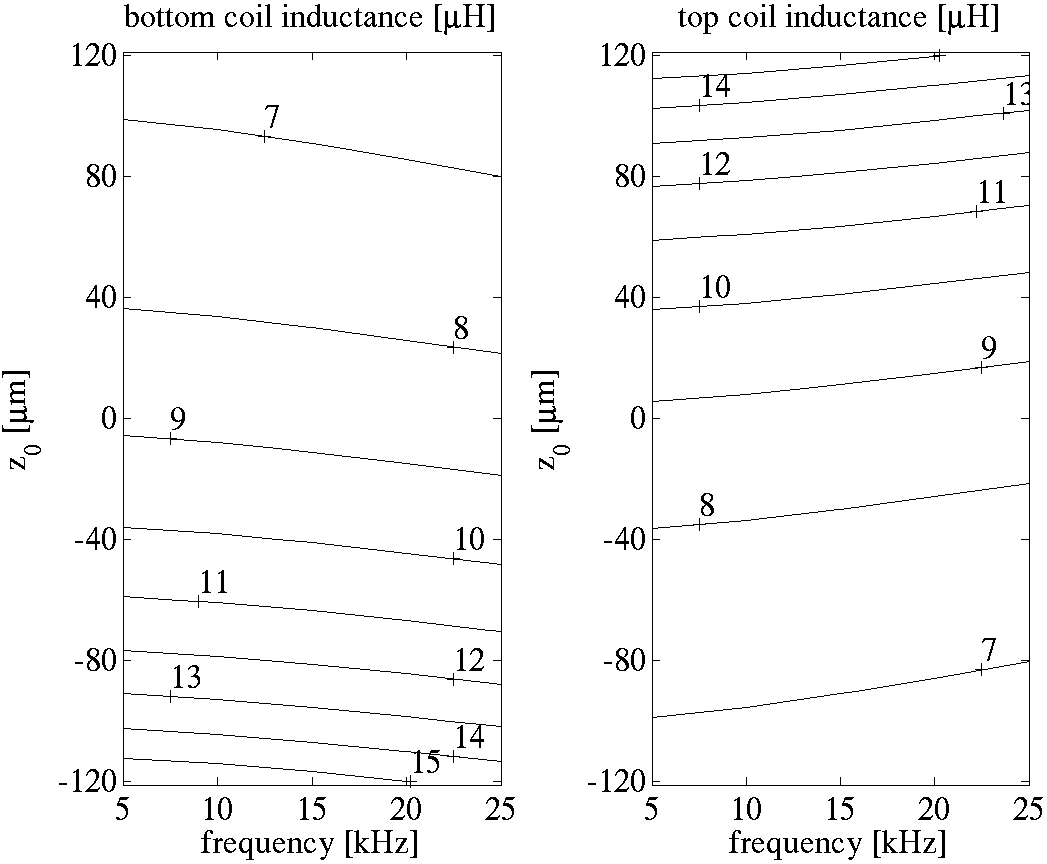}
		\caption{Top and bottom inductance as a function of $z_{0}$.}
		\label{fig:induct}
	\end{center}
\end{figure}

Feeding the coils with sinusoidal currents and running the
solution in the frequency domain at several positions $z_{0}$ of the
mover allows to compute the top and bottom inductances as
a function of the current frequency.
Provided that the current values are small enough, to avoid
an extra power dissipation, and they are phased, to have a null resulting
force on the mover, such a run allows to compute the top and bottom
inductances as a function of $z_{0}$. The results, summarized in
{\mbox{Fig.\ \ref{fig:induct}}}, show that the inductance-position
relationship is robust from the control system standpoint at
all the considered frequencies: over the $\unit[240]{\upmu m}$ $z_{0}$
range the inductance typically changes by a factor of 2.5.

\subsection{TIME-DOMAIN: THE CLOSED LOOP}						\label{sec:time_analysis}

Naming $M = \unit[1.2]{g}$ the mover
mass, $m_{0} = \unit[10]{g}$ the typical payload due to the ASM,
$F$ the force applied to the mover, whose position is $z$, $c$
is the viscous damping coefficient and $K$ is the bending stiffness constant of the
portion of ASM which applies the elastic force $Kz$ to the actuator shaft,
the dynamics of the system is governed by {\mbox{Eq.\ \ref{eq:masspr}} 
\begin{equation}
	F = \frac{d^{2}z}{dt^{2}} + 2 \zeta \omega_{0} \frac{dz}{dt} + \omega_{0}^{2} z	\label{eq:masspr}
\end{equation}
where $ \displaystyle \zeta=\frac{c}{2 \sqrt{K(M+m_{0})}}$
and $ \displaystyle \omega_{0}=\sqrt{\frac{K}{m_{0}+M}}$ 
are the the damping ratio and the natural frequency of the system, respectively.
Adding to the to the FEM described in {\mbox{Sec.\ \ref{sec:fem}}}
the {{\em{ALE}} (Arbitrary Lagrangian-Eulerian
method\cite{comsol_2011}) application
mode, which implements a deformable mesh, the Ordinary Differential
Equation (ODE) defined by {\mbox{Eq.\ \ref{eq:masspr}} allows
to describe the dynamics of the mover.

\subsubsection{The control system}								\label{sec:cs}

The control system, schematized in {\mbox{Fig.\ \ref{fig:cntrl_scheme}}}, is based on
a preshaper and a Proportional-Velocity control. Although the magnetic force
$F$ is strongly non linear both in terms of position $z$ and current $I$,
so that a classic linear controller could not work properly, the reference
signal, the set point $z^{*}$, is always a step of variable amplitude.
Therefore, we can
easily apply an open loop current and desired position signals that drive the actuator
to $z^{*}$ without exciting $\omega_{0}$. In fact, the 2-order system of
{\mbox{Eq.\ \ref{eq:masspr}} may be critically damped simply choosing 
the value of $C$ (the {\em{electronic damping}}) that makes $\zeta=1$.
Defining $t_{s}$ the desired settling time,
replacing the discontinuous {\em{unit step function}} $H(t)$ defined
in {\mbox{Eq.\ \ref{eq:unit}} with
the {\em{smoothed Heaviside step function}} $s(t-t_{s}/2)$
defined in {\mbox{Eq.\ \ref{eq:heavi}}, where 
$p=p(t)$ is a polynomial that makes $s(t-t_{s}/2)$ 
continuous up to its fourth time derivative, the force $F(t)$ is
computed by means of the {\mbox{Eq.\ \ref{eq:masspr}}} and the current $I(t)$
is computed with $I=g(z,F)$, defined in {\mbox{Sec.\ \ref{sec:MS}}}.
As the actuator may follow with some error the position curve, a proportional
control adjusts the current curve with a correction proportional to 
the error $z-z^{*}$ via the constant $K_{p}$.

\begin{eqnarray}
\label{eq:unit}
	   H(t)  & = &  \left\{	\begin{array}{cl}
				   0		& \mbox{if $ t < 0 $} \\ 
				   1		& \mbox{if $ t > 0 $}
				\end{array}
			\right.  
\end{eqnarray}
\begin{eqnarray}
\label{eq:heavi}
	   s(t)  & = &  \left\{	\begin{array}{cl}
				   0		& \mbox{if $ t < -t_{s} $} \\ 
				   p(t)		& \mbox{if $ -t_{s} \leq t \leq t_{s} $} \\
				   1		& \mbox{if $ t > t_{s} $} 
				\end{array}
			\right.  
\end{eqnarray}

\begin{figure}[ht]
	\begin{center}
		\includegraphics[width=.99\textwidth]{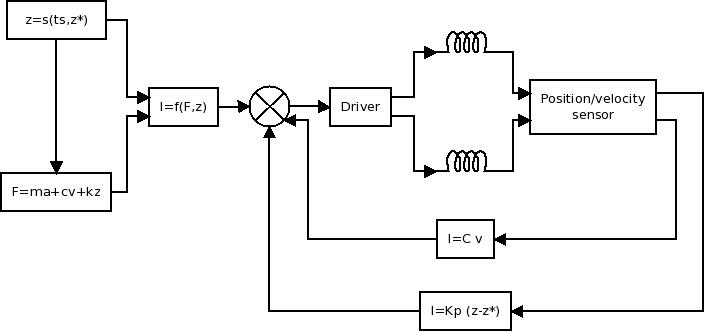}
		\caption{Closed loop scheme. See the text for a discussion.}
		\label{fig:cntrl_scheme}
	\end{center}
\end{figure}

\subsubsection{The electronics}

The pre-shaping defined in {\mbox{Sec.\ \ref{sec:cs}}}
dramatically simplifies the hardware of the
control system, whose output supplies the current generator.
As the inductance depends on the mover position, the electronics
of the current generator, provided with an inductance measure circuit,
works also as a displacement sensor, supplying the control system
with the feed-back values $z$ and $\dot{z}$ with the needed precision.
This capability is provided by two {\em{sniffers}},
one for each coil (included in the ``Position/velocity sensor''
block in {\mbox{Fig.\ \ref{fig:cntrl_scheme}}}),
synchronized with the driver (which embodies the
current generator) via a control logic.
Equipped with a processor that acquires the coil voltage,
the {\em{sniffers}} perform some computational tasks in order to infer
the error displacement $z-z^{*}$, as well as, by a time derivation,
the velocity $v=\dot{z}$.
The driver is a switching-like device, which
controls both the coils. This simple, effective and
low-consumption electronic design, which
adopts the more recent devices developed for power management
applications, such as switching power suppliers, greatly contributes
to reduce the thermal impact.

\subsection{The step response}

In order to test the performances of the control system,
we have selected
$c=0$ and $K=\unit[1]{N \times \upmu m{^{-1}}}$,
a value corresponding to the rigidity of a 
$\unit[1.6]{mm}$ tick Zerodur ASM whose actuators are separated by
$\unit[25]{mm}$, in {\mbox{Eq.\ \ref{eq:masspr}}} and
$K_{p}=\unit[2 \times 10^{8}]{A \times m{^{-1}}}$,
$C=\unit[126.5]{kg \times s{^{-1}}}$,
and $t_{s}=\unit[.5]{ms}$
in {\mbox{Fig.\ \ref{fig:cntrl_scheme}}}.
Commanding $\unit[\updelta = 1 \; \mathrm{to} \; 5]{\upmu m}$, where
$\delta=z-z_{0}$ is the requested stroke,
for $z_{0}$ ranging from from $-100$ to $\unit[100]{\upmu m}$,
gives very good time responses: the settling time ranges from
$\unit[.35 \; \mathrm{to} \; .37]{ms}$, largely
below the specification in terms of bandwidth.
{\mbox{Fig.\ \ref{fig:cl_all}}} shows the step response 
for the lower and upper limits of the considered $\delta$ values.

\begin{figure}[H]
	\begin{center}
		\begin{tabular}{cc}
			\includegraphics[width=.47\textwidth]{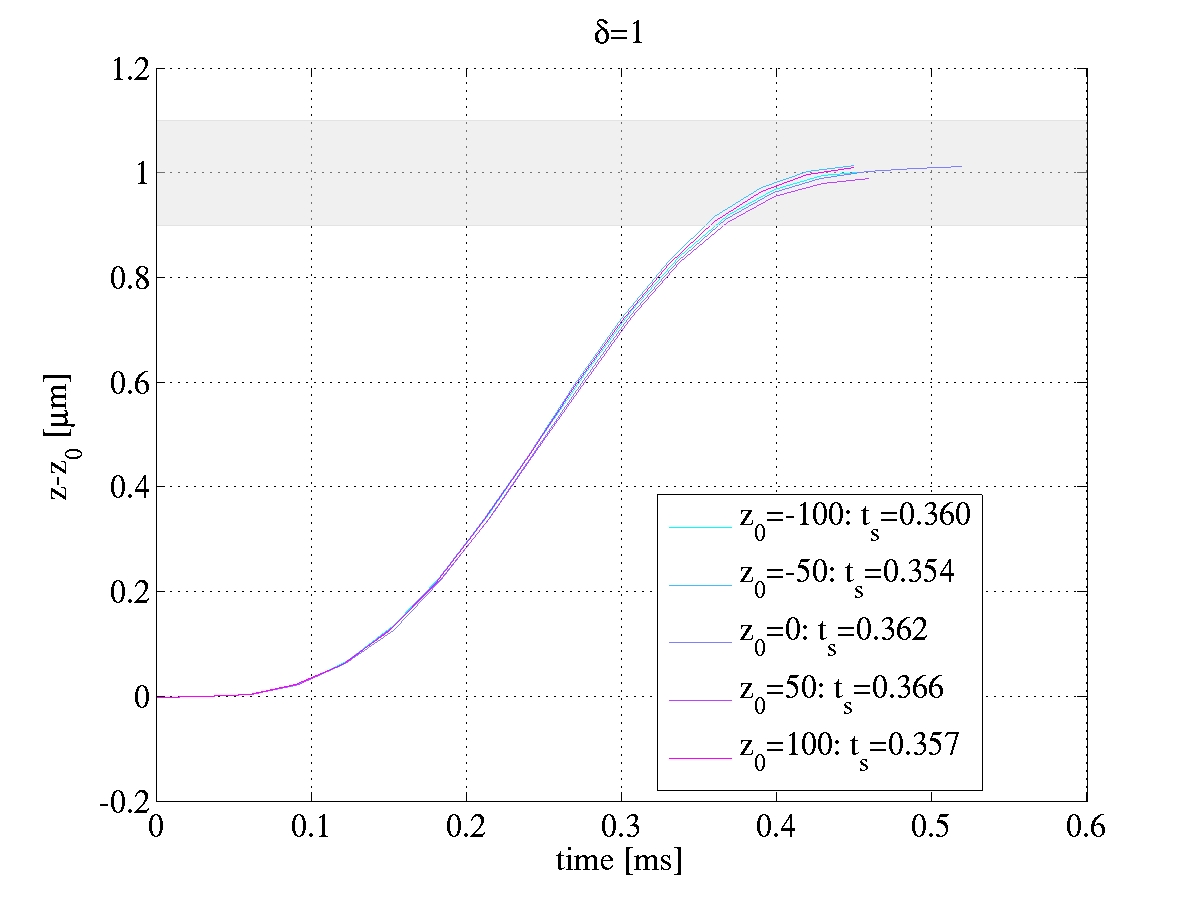} & \includegraphics[width=.47\textwidth]{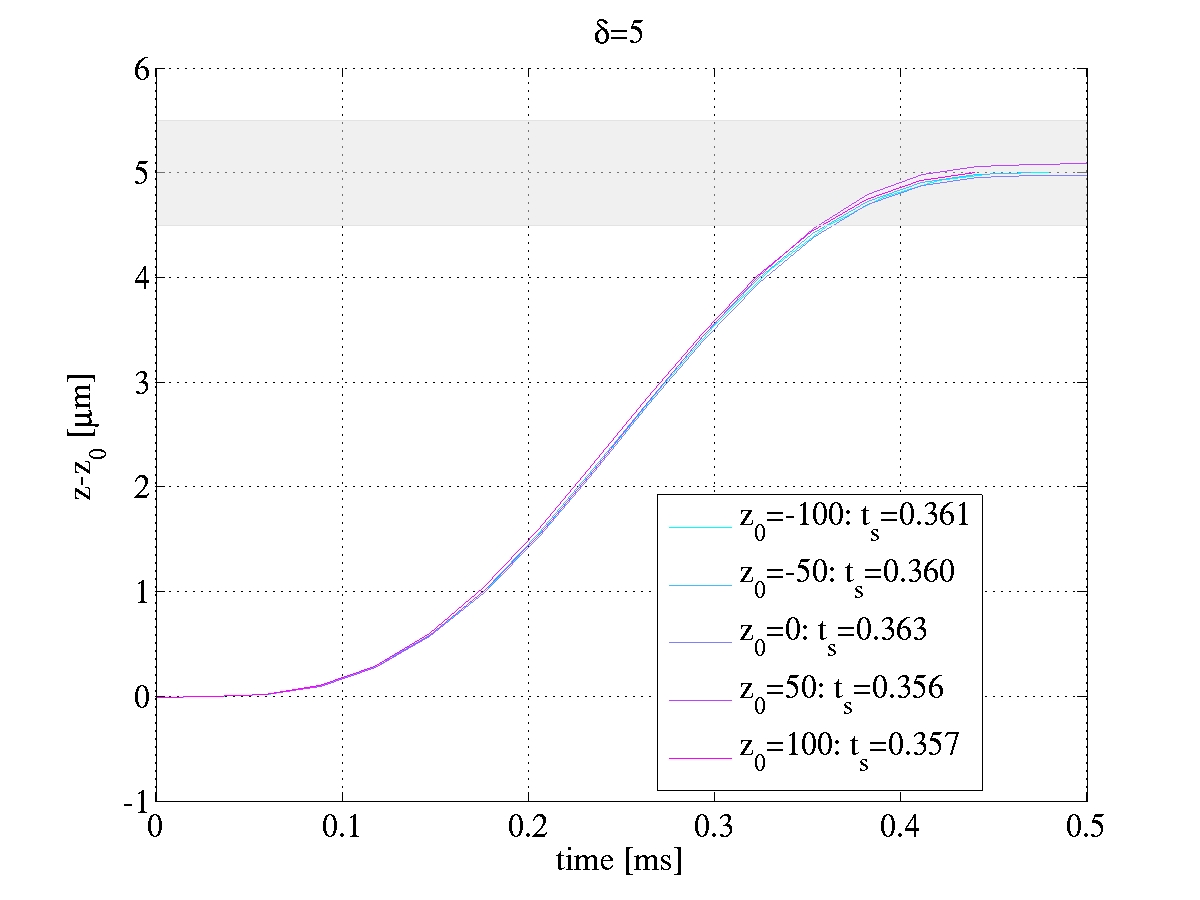} \\
		\end{tabular}
	\end{center}
	\caption[clall]{\label{fig:cl_all}Step response
	for $\unit[100 \leq z_{0} \leq 100]{\upmu m}$
	and
	$\unit[\delta = 1 \; \mathrm{and} \; 5]{\upmu m}$.
	The shaded strips indicate the domains where $ \left| (z-\delta)/\delta \right| \leq 10\%$.
	}
\end{figure}

\section{EXPERIMENTAL VALIDATION}								\label{sec:test}

\begin{figure}[H]
	\begin{center}
		\subfigure[]{\label{fig:valid-graph}\includegraphics[width=.495\textwidth]{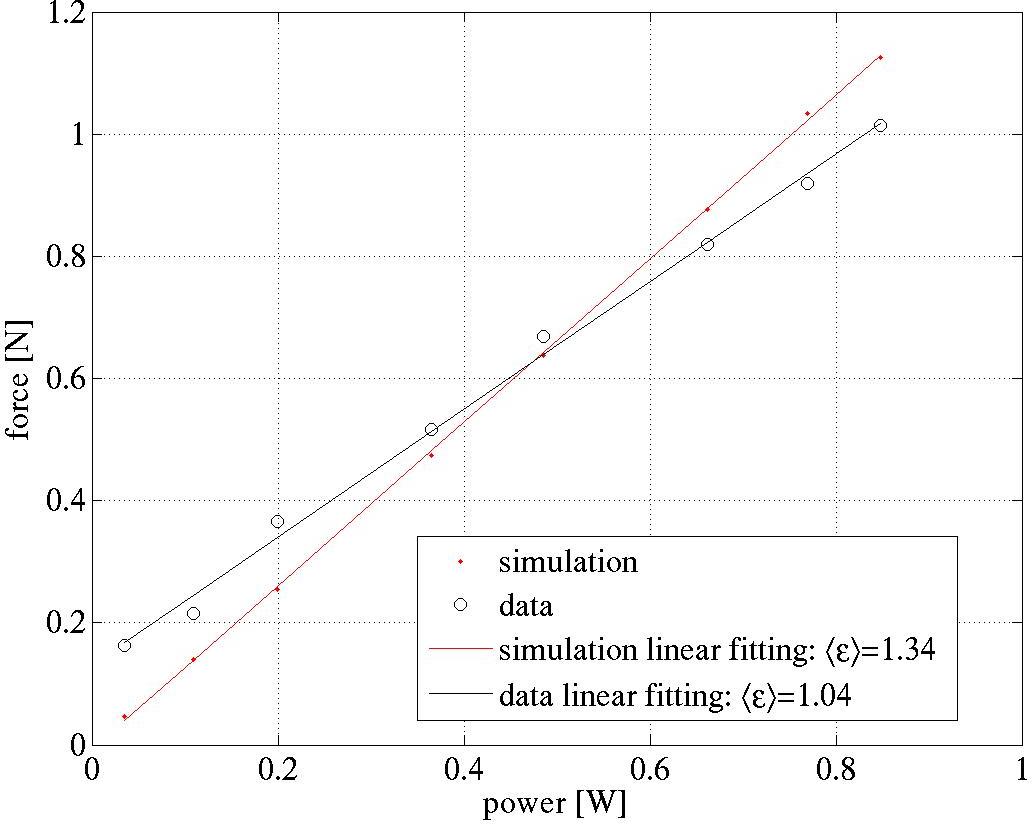}}
		\vspace{-5mm}
		\subfigure[]{\label{fig:valid-pict}\includegraphics[width=.495\textwidth]{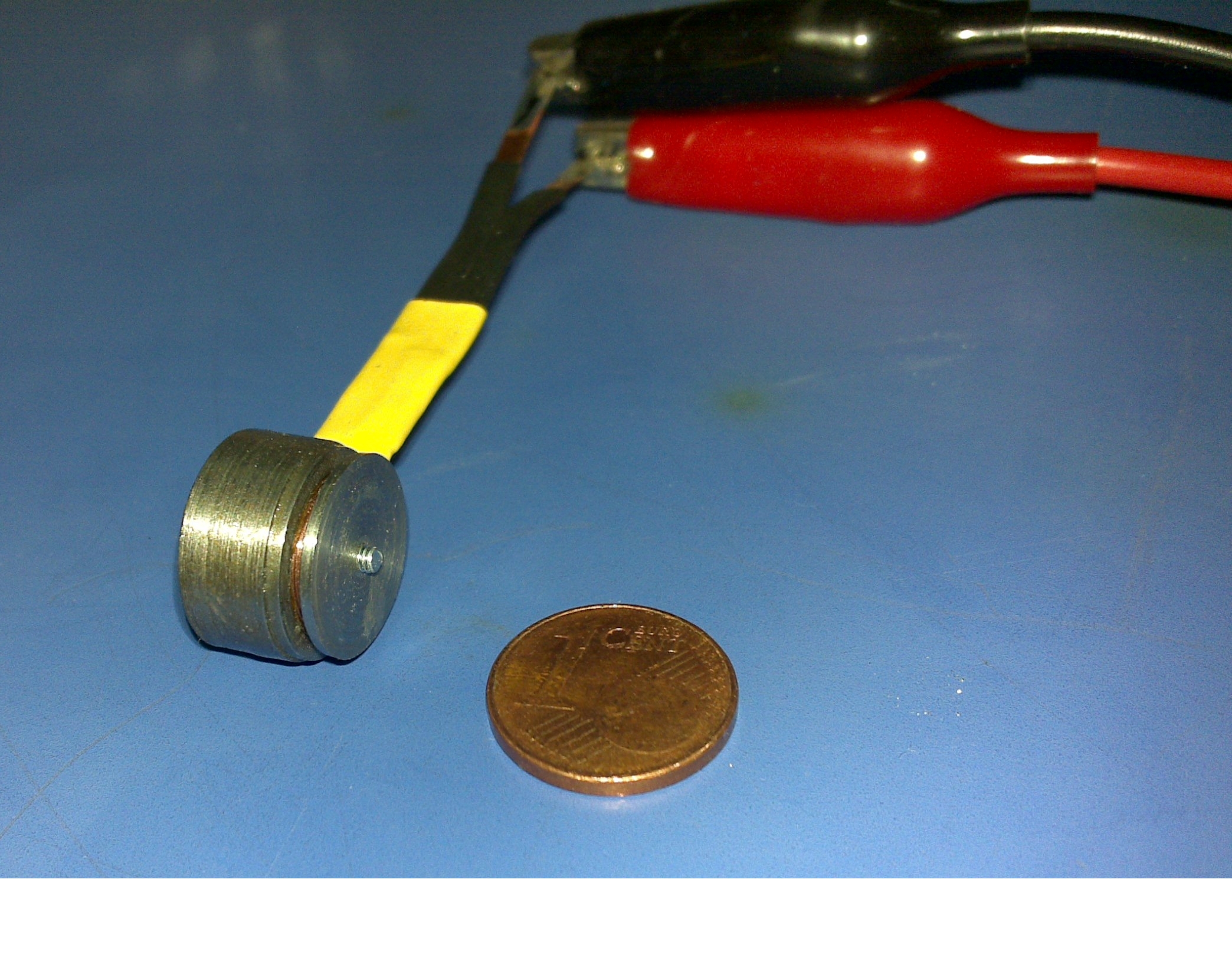}}
	\end{center}
	\caption{Force $F$ as a function of the DC power \subref{fig:valid-graph} of the prototype \subref{fig:valid-pict}. See the text for a discussion.}
	\label{fig:valid}
\end{figure}
A preliminary, magnetostatic only test has been run in order to 
verify the reliability of the FEA model. Instead of building
an expensive and delicate complete actuator
with a Somaloy core, a simpler single stator/mover unit, shown in
{\mbox{Fig.\ \ref{fig:valid-pict}}}, made of the
cheaper C40 soft iron, have been manufactured in the laboratory.
The nominal $\unit[200]{\upmu m}$ air gap has been replaced
by a $\unit[330]{\upmu m}$ tick plastic spacer.
The results of the magnetostatic solution of the FEM model
of such a prototype, taking as B-H curve some data measured on
a sample of the C40, are shown in {\mbox{Fig.\ \ref{fig:valid-graph}}},
along with the experimental data.
The difference between the values computed by the simulation and
the measured ones ranges from $8\%$ to $11\%$ in the current
span $\unit[3 \; \mathrm{to} \; 4.7]{A}$~---~corresponding to
$\unit[365 \; \mathrm{to} \; 848]{mW}$. We note that the efficiency,
also reported in {\mbox{Fig.\ \ref{fig:valid-graph}}},
is reduced by a factor of four with respect to the full model described
in the previous sections, mainly because of the $65\%$ increasing
of the gap. Considering that the available C40 data 
were pretty rough and the manufacturing of the prototype
was not as accurate as the specifications require, the numerical model
is magneto-statically sound.

\section{CONCLUSIONS AND FUTURE WORK}								\label{sec:concl}

Based on a simple and very effective magnetic circuit and on a highly
performing control system, VRALA allows to implement the AO 
technology at visible wavelengths, where large forces and unprecedented
actuator densities are required.

The FEA multiphysics simulation results demonstrate that this actuator exhibits
very low power dissipations and applies the corrections with an excellent bandwidth,
thanks to the hardware and the software designs of the control system. With
an efficiency of $\unit[4.65]{N \times W^{-1}}$ and an overall
radius that allows actuator separations up to $\unit[25]{mm}$, VRALA can
provide strokes up to $\unit[5]{\upmu m}$ in less than
$\unit[.37]{ms}$~---~largely below the bandwidth goal of
$\unit[2]{kHz}$~---~with an extremely low thermal impact.
The compact, effective and simple control system exhibits
the capability of driving the actuator while sensing its motion without
any additional feed-back component.

The force measures carried out on the preliminary prototype, although built with
a rough, off-the-shelf available material, show that the magnetostatic
computations are correct.

These promising results indicate the future developments, in terms
of simulation and laboratory activities.

Further, more complex numerical simulations are needed:
complete 2D and a 3D FEA models must be
developed in order to simulate the complex response of the system, in
terms of mechanical and thermal deformations, possible electromagnetic
variations caused by tolerances and mutual effect by close actuators,
as well as the fluid dynamics governing the convective heat exchanges.

A full Somaloy prototype is planned,
in order to verify the response of the material selected to build the
iron core. Because of the efficiency is proportional to the filling
factor, replacing the conventional solenoid with a pre-wounded copper strip 
is also planned, in order to increase, according to some preliminary
computations, the filling factor up to $\unit[\approx .9]{}$.
The electronics of the control system, currently under construction,
will allow to test the real dynamics of the actuator when operating
in turbulence corrections.
Finally, the construction of 4-by-4 actuators demonstrator will
provide some important data on the actual opto-mechanical response of
the full system and
the multi-channel control electronics, currently in the feasibility
phase.

\acknowledgments     

This study was supported by the TECNO INAF 2009 grant from the
Italian Istituto Nazionale di Astrofisica (INAF).
The authors would like to thank Riccardo Crosa
and Federico Della Ricca from Höganäs Italy,
for insight and guidance through the Somaloy materials.


\bibliography{citat8}		
\bibliographystyle{spiebib}	

\end{document}